\documentclass[showpacs]{revtex4}

\usepackage{graphicx}
\usepackage{dcolumn}
\usepackage{amsmath}
\makeatletter
\def\btt#1{\texttt{\@backslashchar#1}}
\DeclareRobustCommand\bblash{\btt{\@backslashchar}}
\makeatother

\begin{document}

\title{Non-spherical collapse of a two fluid star}

\author{S. G.~Ghosh}
\email{sgghosh@iucaa.ernet.in} \affiliation{BITS, Pilani DUBAI
Campus, P.B. 500022, Knowledge Village, DUBAI, UAE  and
Inter-University Center for Astronomy and Astrophysics,
 Post Bag 4 Ganeshkhind,  Pune - 411 007, INDIA }

\date{\today}

\begin{abstract}
We obtain the analogue of collapsing Vaidya-like solution to include
both a null fluid and a string fluid, with a linear equation of
state ($p_{\bot } = k \rho$), in non-spherical (plane symmetric and
cylindrically symmetric) anti-de Sitter space-timess. It turns out
that the non-spherical collapse of two fluid in anti-de Sitter
space-times, in accordance with cosmic censorship, proceed to form
black holes, i.e., on naked singularity ever forms,  violating hoop
conjecture.
\end{abstract}


\maketitle

\section{Introduction}
 It is well known that the general relativity admits solutions with
 singularity, and that such solutions can be produced by
 non-singular initial data.  The singularity theorems \cite{he} revealed
that the occurrence of singularities is a generic property of
space-times in classical general relativity (GR).  These theorems
however can not predict  the final state of such a singularity.
However, there is common belief that a cosmic censor exists who
safely hides the singularity inside a black hole, actually this is a
famous cosmic censorship conjecture (CCC), first formulated by
Penrose \cite{rp}. The CCC remains as one of the most outstanding
unresolved question in GR. However, there are many known examples in
the literature showing that both naked singularities and black holes
can form in gravitational collapse \cite{r1}. The central shell
focusing singularity can be naked or covered depending upon the
choice of initial data. There is a critical branch of solution where
a transition from naked singularity to black hole occurs. In
particular gravitational collapse of spherical matter in the form of
radiation (null fluid) described by Vaidya metric \cite{pc} is well
studied for investigating CCC in four dimension \cite{vc} as well as
in higher dimensions \cite{gd}.  These counterexamples are
restricted to spherical symmetry.  Are these singularities an
accident of spherical symmetry?

For the non-spherical collapse, Thorne \cite{ks} has proposed hoop
conjecture: that collapse will yield a black hole only if a mass $M$
is compressed to a region with circumference $C \leq 4 \pi M$ in all
directions. If hoop conjecture is true, naked singularities may form
if collapse can yield  $C \geq 4 \pi M$ in some direction.   Thus,
planar or cylindrical matter will not form a black hole (black plane
or black string) \cite{ks}. Indeed, Shapiro and Teukolsky \cite{st}
showed that collapse of a prolate spheroid leads to a spindle
singularity without horizon, i.e., a naked singularity may form in
non-spherical relativistic collapse. Also, hoop conjecture was given
for space-times with a zero cosmological term and in the presence of
negative cosmological term one can expect the occurrence of major
changes. It is clear that, in spherically symmetric, the effect of
adding negative cosmological term does not alter the final fate of
collapse \cite{wm}. However, Lemos \cite{jpl} has shown that planar
or cylindrical black holes form rather than naked singularity from
gravitational collapse of a planar or cylindrical matter
distribution (null fluid) in an anti-de Sitter space-time, violating
in this way the hoop conjecture but not CCC.  Here the negative
cosmological constant plays crucial role as in the BTZ black holes.

Several generalization of the Vaidya solution in which the source is
a mixture of fluids and radiation have been obtained
\cite{vc1,ww,vh,gd1}. Recently, Glass and Krisch  \cite{gk} came up
with a generalized Vaidya solution for a two fluids: a null fluid
with a string fluid. They show that by allowing the Swarzschild mass
as a function of retarded time creates an atmosphere for two fluid:
Vaidya radiation fluid in addition to a string fluid.  The solution
is very important in view of recent links between black holes and
string theories \cite{mc,sk,fl}.  Further, the string is very
important ingredient in many physical theories and idea of string is
fundamental to superstring theories \cite{as}. The solution has been
employed to look into the consequence of string fluid on the
formation of naked singularities in Vaidya collapse \cite{gg,gs}.
The effect of string is a shrinkage of the naked singularity initial
data space, or an enlargement of the black hole initial data space
of Vaidya collapse \cite{gs}.

In this paper, we shall first obtain the general solution for a
two-fluid: a null fluid and a string fluid, with the equation of
state ($p_{\bot}  = k \rho$ and $\rho=\rho_0$), in non-spherical
(plane symmetric and cylindrical symmetric) anti-de Sitter
space-times. We then see how the results that were presented in
\cite{jpl} get modified in the presence of the string fluid.

We find that the non-spherical collapse, in contrast to spherically
symmetric collapse where naked singularity inventible, lead to black
holes, an accordance with CCC, giving explicit counterexample to
hoop conjecture.
\section{Non-spherical two fluid model}
In this section , we extend the Glass and Krisch  solutions
\cite{gk}, to non-spherical (plane symmetric and cylindrical
symmetric) anti-de Sitter space-times. Let us first consider the
case of plane symmetry. The metric of general plane symmetric
space-time, expressed in terms of Eddington advanced time coordinate
(outgoing coordinate) $v$, reads:
\begin{equation}
ds^2 = -e^{\psi(v,r)}dv \left[e^{\psi(v,r)} f(v,r) dv
 +  2 dr \right] + r^2 \alpha^2 (dx^2+ dy^2), \label{eq:me}
\end{equation}
where  $- \infty \leq x, \: y \leq \infty$ are coordinate which
describe two-dimensional zero-curvature space which has topology
$R\times R$,  $- \infty \leq v  \leq \infty$ is null coordinate
called the retarded time, and  $0  \leq r \leq \infty$ is the radial
coordinate. Further, $e^{\psi(v,r)}$ is an arbitrary function and
where $3 \alpha^2 = - \Lambda > 0$ denote negative cosmological
constant. It is useful to introduce a local mass function $m(v,r)$
defined by $f = 1 - {2 m(v,r)}/{r}$. For $f = m(v)/r$ and $\psi=0$,
the metric (\ref{eq:me}) reduces to the plane symmetric Vaidya-like
metric \cite{bk,cs}. Initially $f= M_0/r$ (with $\psi=0$) provides
the vacuum Taub solution \cite{at}.

It is the field equation $G^0_1 = 0$ that leads to $ e^{\psi(v,r)} =
g(v)$. However, by introducing another null coordinate $
e^{\psi(v,r)} = g(v)$, we can always set without the loss of
generality, $\psi(v,r) =0$. Hence, the metric takes the form:
\begin{equation}
ds^2 =  -\left[1 - \frac{2 m(v,r)}{r}\right] d v^2 + 2 d v d r +
\alpha^2 r^2 (dx^2+ dy^2). \label{eq:me1}
\end{equation}
The use of a Newman-Penrose null tetrad formulism leads to Einstein
tensor of the form \cite{gk,mc}:
\begin{equation}
G_{ab} = - 2 \Psi_{11}(l_a n_b + l_b n_a + m_a \overline{m}_b +
\overline{m}_a m_b) - 2 \Psi_{11} l_a n_b - 6\Lambda.. \label{ee}
\end{equation}
Here the null tetrad Ricci scalars are
\begin{eqnarray}
  \Psi_{11} = \frac{1}{r^2} \left[{2 \frac{\partial m}{\partial r} - r \frac{\partial^2 m}{\partial r^2} -
  1}\right] \\
  \Psi_{22} = - \frac{1}{r^2} \frac{\partial m}{\partial v} \\
  \Lambda = \frac{1}{r^2} \left[{2 r \frac{\partial^2 m}{\partial r^2} - 2 \frac{\partial m}{\partial r} -
  1}\right]
\end{eqnarray} the
principal null geodesic vectors are $l_a,n_a$ of the form
\begin{equation}
l_a=\delta _a^u,\qquad n_a=f/2\delta _a^u - \delta _a^r,
\end{equation}
where $l_al^a=n_an^a=0,$ $l_an^a=-1$.  The metric (\ref{eq:me1})
admits an orthonormal basis defined by four unit vectors
\begin{eqnarray}
\hat{u}_a=f^{1/2}\delta _a^u - f^{-1/2}\delta _a^r,\qquad \hat{r}%
_a=f^{-1/2}\delta _a^r, \\
\hat{\mathbf{x}}_a= \alpha x\delta _a^x ,\qquad \hat{%
\mathbf{y}}_a= \alpha y \delta _a^y ,  \label{ov}
\end{eqnarray}
where $\hat{u_a}$ is a timelike unit vector and $\hat{r_a}$,
$\hat{\mathbf{x}_a}$, $\hat{\mathbf{y}_b}$ are unit space-like
vector such that
\begin{equation}
g_{ab} = \hat{u_a}\hat{u_b} - \hat{r_a}\hat{r_b} -
\hat{\mathbf{x}_a}\hat{\mathbf{x}_b}- \hat{\mathbf{y}_a}
\hat{\mathbf{y}_b}, \label{mt1}
\end{equation}

Associated with the string world sheet we have the string bivector
defined by
\begin{equation}
\Sigma^{ab}=\epsilon^{AB} \frac{dx^{a}}{dk^A}\frac{dx^{b}}{dk^B},
\label{sb}
\end{equation}
where $\epsilon^{AB}$ is two dimensional Levi-Civita symbol.  It is
useful to write the bivector, in terms of the unit vectors, as
\begin{equation}
\Sigma ^{ab}=\hat{r}^a\hat{u}^b-\hat{u}^a\hat{r}^b,  \label{bv}
\end{equation}
and the condition that the worldsheet are timelike, i.e., $\gamma
=1/2\Sigma ^{ab}\Sigma _{ab}<0$ implies that only the $\Sigma ^{ur}$
component is non-zero, therefore one obtains:
\begin{equation}
\Sigma ^{ac}\Sigma _c^b=\hat{u}^a\hat{u}^b-\hat{r}^a\hat{r}^b.
\end{equation}
The string energy-momentum tensor for a cloud of string, by analogy
with the one for the perfect fluid, is written as \cite{gk,mc}:
\begin{equation}
T_{ab}^{\left( s\right) }=\rho \Sigma _a^c\Sigma _{cb}-p_{\bot
}h_{ab}, \label{ems}
\end{equation}
The energy-momentum of two fluid system is $T_{ab} = T^{(n)}_{ab} +
T^{(s)}_{ab}$, where
\begin{equation} T_{ab}^{(n)}=\psi l_al_b.
\end{equation}
It is the null fluid tensor corresponding to the component of the
matter field that moves along the null hypersurfaces
$v=\mbox{const.}$ The effective energy momentum tensor for two fluid
system, in terms of the unit vectors, can be cast as:
\begin{equation}
T_{ab}=\psi l_al_b + \rho
\hat{u}_a\hat{u}_b+p_r\hat{r}_a\hat{r}_b+p_{\bot }\left(
\hat{\mathbf{x}}_a\hat{\mathbf{x}}_b+\hat{\mathbf{y}}_a\hat{\mathbf{y}}_b\right).
\label{emsv}
\end{equation}
For $\rho = p_r = p_{\bot }= 0$, Eq.~(\ref{emsv}) reduces to
stress-energy tensor which gives Vaidya metric \cite{bk,cs}.
Comparing Eq.~(\ref{ee}) and Eq.~(\ref{emsv}), the Einstein field
equations now take the form:
\begin{eqnarray}
\psi =\frac{1}{4 \pi r^2} \frac{\partial m}{\partial v}, \\
\rho =-p_r= - \frac{1}{8\pi r^2}+\frac{1}{4\pi r^2}\frac{\partial
m}{\partial r} + \frac{1}{8\pi } 3 \alpha^2, \\
p_{\bot }= - \frac{1}{8\pi r} \frac{\partial^2 m}{\partial r^2} -
\frac{1}{8\pi } 3 \alpha^2.
\end{eqnarray}
Next, we construct a model by assuming the tangential pressure
$P_{\bot}$ to energy density $\rho$ by a linear equation of state
$p_{\bot } = k \rho$ \cite{tplw}, which leads to
\begin{equation}
\frac{\partial^2 m}{\partial r^2} + \frac{2k}{r} \frac{\partial
m}{\partial r} + 3 \alpha^2 (1+k) = \frac{k}{r}. \label{de}
\end{equation}
The integration of the this equation reads:

\begin{equation}
m(v,r) = \left\{ \begin{array}{ll}
       {r}/{2}+ M(v) + \frac{S(v)r^{1-2k}}{1-2k} - {\alpha^2
r^3}/{2},                   &   \hspace{0.05in}    \mbox{if $k \neq 1/2$}, \\
         & \\

       {r}/{2}+ M(v) + S(v) \ln r - {\alpha^2
r^3}/{2} & \hspace{0.05in}      \mbox{if $k = 1/2$}.
                \end{array}
        \right.                         \label{eq:mv}
\end{equation}
where $M(v)$ and $S(v)$ are arbitrary functions of integration and
may not be related, and $M(u)$ can be treated as Vaidya-mass and
$S(v)$ as contribution from string fluid.  The class of solution
discussed above, in general, belongs to Type II fluid defined in
\cite{he}. When $m=m(r)$, we have $\psi$=0, and the matter field
degenerates to type I fluid \cite{ww}. In the rest frame associated
with the observer, the energy-density of the matter will be given
by,
\begin{equation}
\psi = T^r_v,\hspace{.1in}\rho = - T^t_t = - T^r_r , \label{energy}
\end{equation}
 and the principal pressures are $P_i =
T^i_i$ (no sum convention).  Therefore $P_r = T^r_r = - \rho$ and
$P_{x} = P_{y} = k \rho$. \\

\noindent \emph{The weak energy conditions} (WEC): The energy
momentum tensor obeys inequality $T_{ab}w^a w^b \geq 0$ for any
timelike vector, i.e.,
\begin{equation}
\psi \geq 0,\hspace{0.1 in}\rho \geq 0,\hspace{0.1 in} P_{x} \geq 0,
\hspace{0.1 in} P_{y} \geq 0. \label{wec}
\end{equation}
However, $\psi
> 0$ gives the restriction on the choice of the functions $M(v)$
and $S(v)$. We observe $\psi > 0$ requires,
\begin{equation}
\frac{\partial M}{\partial v} + \frac{\partial S}{\partial
v}\hspace{.01in}\frac{r^{1-2k}}{1-2k} > 0.\end{equation} This, in
general, is satisfied as long as both ${\partial M}/{\partial v}$
and $ {\partial S}/{\partial v}$ are greater than zero. Furthermore,
to guarantee the positivity of the energy density, we must have $k <
1/2$ and $S(v) > 0$, or $k > 1/2$ and $S(v) < 0.$  For $k=1/2$, weak
energy condition can not be satisfied for all $r$ \cite{vh}.
Therefore, we shall not consider this case further.

In summary, we have show that the metric:
\begin{eqnarray}
ds^2 =  - \left(\alpha^2 r^2 -  \frac{2 M(v)}{r} - \frac{2
S(v)}{(1-2k)r^{2k}} \right) dv^2 + 2 dv dr  + \alpha^2 r^2
(dx^2+dy^2). \label{eq:me2}
\end{eqnarray}
is solution of the Einstein equations for the energy momentum tensor
(\ref{emsv}) with equation of state $p_{\bot } = k \rho$.
 Thus we have extended the collapsing plane symmetric Vaidya-like
\cite{bk,cs} to include both a null fluid and a string fluid and we
shall call it generalized plane symmetric Vaidya metric.  The metric
(\ref{emsv}) is solution of Einstein equation for two fluid system:
a null fluid and a string fluid in plane symmetric anti-de Sitter
space-times. This metric was also obtained by Cai {it et al.}
\cite{cz} by treating the null fluid to behave like perfect fluid.
The physical quantities for this metric are given by
\begin{equation}
\psi = \frac{1}{4\pi r^2} \left(\frac{\partial M}{\partial
v}+\frac{\partial S}{\partial v} \frac{r^{1-2k}}{1-2k} \right)
\label{eq:psi}
\end{equation}
\begin{equation}
\rho = \frac{1}{4\pi} \frac{S(v)}{r^{2k}}  = - p_r \label{eq:rho}
\end{equation}
\begin{equation}
p_{\bot } = k \rho \label{pt}
\end{equation}
There are special cases of this solution which are already known:
One is plane symmetric anti-de Sitter Vaidya metric \cite{lv}, which
arises for $S(v)=0$ (vanishing $\rho$ and $p$). When $k=1$ and
$2S(v)= - Q^2(v) $, the solution (\ref{eq:me2}) reduces to plane
symmetric Bonnor-Vaidya anti-de Sitter metric \cite{cz}. The Riemann
anti-de Sitter metric arises when, in addition to $S(v)=0$, $M(v) =
M_0$ is a constant. The vacuum background (M=S=0) to
Eq.(\ref{eq:me2}) is anti-de Sitter space-time.

\noindent {\emph{Horizons:}}As demonstrated by York \cite{jy},
horizons can be obtained by noting that (i) apparent horizons are
defined as surface such that $\Theta \simeq 0$ and (ii) event
horizons are surfaces such that $d \Theta /dv \simeq 0$, where
$\Theta$ is expansion.  For the metric (\ref{eq:me2}), we have
\begin{equation}\label{ah1}
\Theta= \frac{1}{r} \left[\alpha^2 r^2 - \frac{2 M(v)}{r} - \frac{2
S(v)}{(1-2k)r^{2k}} \right],
\end{equation}
 Since the York conditions require that at apparent horizons $ \Theta$
vanish, it follows form the Eq.~(\ref{ah1}) that apparent horizons
will satisfy
\begin{equation}\label{ah2}
\alpha^2 r^3 - 2 S \frac{r^{1 - 2k}}{1 - 2k} - 2M=0,
\end{equation}
which in general has positive solutions and hence final
configuration would be of multiple horizon depending on the value of
$k$.

{\it {Cylindrically symmetric space-time}:} We turn our attention to
the cylindrical symmetric ant-de Sitter space-time. The metric
(\ref{eq:me1}) in the Cylindrical space-time has the form:
\begin{equation}
ds^2 = - \left(\alpha^2 r^2 -  \frac{2 M(v)}{r} -
\frac{S(v)}{(1-2k)r^{2k}} \right) dv^2 + 2 dv dr +  r^2 d\theta^2 +
\alpha^2 r^2 dz^2, \label{eq:sme}
\end{equation}
where $ \infty < v$, $z < \infty $, $0 \leq r < \infty $ and $0 \leq
\theta \leq  2\pi$. The two dimensional surface has topology of $R
\times S^1$.  The analysis given above, with suitable modifications,
is valid in  the cylindrical anti-de Sitter space-time as well.
Hence, to conserve space, we avoid repetition of the detailed
analysis (being similar to plane symmetric case).  Hence, we shall
limit ourself to plane symmetric space-time in the further analysis.
\section{Gravitational collapse}
To determine the solution completely, the boundary condition must be
imposed, as well as the specification of the space-time in the
vacuum region exterior to star. The initial radius of star from
which collapse begin would be given by $p_r=0$ which would demand
that the function $S(v)=0$.  The Kretschmann scalar ($K = R_{abcd}
R^{abcd}$, $R_{abcd}$ is the Riemann tensor) for the metric
(\ref{eq:me}) reduces to
\begin{equation}
\mbox{K} = \frac{48}{r^6} M^2(v) + \frac{16 K_1}{r^{4k+4}}S^2(v) +
\frac{32K_2}{r^{5+2k}}M(v) S(v) +\frac{16 \alpha^2 K_3}{r^{2k+2}}
S(v) + 24 \alpha^4 \label{eq:ks}.
\end{equation}
Here, $K_1 = (4k^4+4k^3+5k^2+1)/(2k-1)^2$, $K_2 =
(-4k^3-4k^2+k+1)/(2k-1)^2$ and $K_3 = (4k^3-8k^2+5k-1) $. So the
Kretschmann scalar diverges only on the plane $r = 0$ for $M$ and $S
\neq 0$, establishing that metric (\ref{eq:me}) is scalar polynomial
singular \cite{he}.

The physical situation is that of a radial influx of two fluid in
the region of the anti-de Sitter universe.  The first shell arrives
at $r=0$ at time $v=0$ and the final at $v=T$. A central singularity
of growing mass developed at $r=0$.
  For $ v < 0$ we have $M(v)\;=\;S(v)\;=\;0$, i.e., the anti-de Sitter metric,
 and for $ v > T$,
$\dot{M}(v)\;=\;{S}(v)\;=\;0$, $M(v)$ is positive definite. The
metric for $v=0$ to $v=T$ is  generalized plane symmetric
Vaidya-anti-de Sitter, and for $v>T$ we have the exterior background
space-time, called Reimann anti-de Sitter solution \cite{lv}. Thus
the metric (\ref{eq:me}) is not asymptotically flat.

Radial ($ \theta$ and $ \phi \,=\,const$.) null geodesics of the
metric (1) must satisfy the null condition
\begin{equation}
\frac{dr}{dv} = \frac{1}{2} \left[\alpha^2 r^2 - \frac{2 M(v)}{r} -
\frac{2 S(v)}{r^{2k}} \right].         \label{eq:de1}
\end{equation}
 The nature (a naked singularity or a black hole) of the
collapsing solutions can be characterized by the existence of radial
null geodesics coming out from the singularity.  The nature of the
singularity can be analyzed by techniques in \cite{r1}. In order to
get an analytical solution, we choose
\begin{equation}
M(v) = \lambda v \; (\lambda > 0) \; \mbox{and} \; S(v) = \mu^2
v^{2k} (\mu^2
> 0) \label{eq:mv1}
\end{equation}
for $0 \leq v \leq T$ \cite{ps}. Let $y \equiv v/r$ be the tangent
to a possible outgoing geodesic from the singularity. In order to
determine the nature of the limiting value of $y$ at $r=0$, $v=0$ on
a singular geodesic, we let
\begin{equation}
y_{0} = \lim_{r \rightarrow 0 \; v\rightarrow 0} y =
\lim_{r\rightarrow 0 \; v\rightarrow 0} \frac{v}{r}.
\label{eq:lm1}
\end{equation}
Using Eqs. (\ref{eq:de1}), (\ref{eq:mv})  and L'H\^{o}pital's rule
we get
\begin{equation}
y_{0} = \lim_{r\rightarrow 0 \; v\rightarrow 0} y =
\lim_{r\rightarrow 0 \; v\rightarrow 0} \frac{v}{r}=
\lim_{r\rightarrow 0 \; v\rightarrow 0} \frac{dv}{dr} =
\frac{2}{\alpha^2 r^2 - 2 \lambda y_{0} - \mu^{2} y_{0}^{2k}}
\label{eq:lm2}
\end{equation}
which implies,
\begin{equation}
 \mu^{2} y_{0}^{2k+1} + 2 \lambda y_{0}^2  + 2 = 0.       \label{eq:ae}
\end{equation}
If Eq. (\ref{eq:ae})
 admits one or more positive real roots then the  central shell focusing
singularity is at least locally naked.  In the absence of positive
roots  of (\ref{eq:ae}), the central singularity is not naked
because in that case there are no outgoing future directed null
geodesics from the singularity (for more details, see \cite{r1}).
Since there are no positive roots to (\ref{eq:ae}), for all
$\lambda$ and $\mu$, the collapse will always lead to a black hole,
i.e., no radial future null geodesics terminate at the singularity.
This is contrast to spherically symmetric collapse where a naked
singularity is inventible \cite{gd}.

 Similar to the above discussion is also valid in
Cylindrical case as well.  Therefore, we can conclude that the
gravitational collapse in the cylindrical symmetric anti-de Sitter
space-time also forms a black hole.
\section{Constant density string fluid ($\rho = \rho_0$ )}
Next. we consider that the string fluid is of constant density ($
\rho_0$) Although strictly constant density is not completely
realistic, the particular analytic solution to Einstein field
equation with this equation of state has provided some insights
concerning stars in general relativity.  For example, the star
represented by this solution has property that it cannot remain in
equilibrium if the size-to-mass ratio is less than 9/4 \cite{ck}. To
obtain the constant density solution, we take $\rho = \rho_0$. With
this it can bee seen that (\ref{eq:de2}) can be easily integrated to
yield:
\begin{equation}
ds^2 = - \left[ \beta^2 r^2 -  \frac{2 M(v)}{r} \right] dv^2 + 2 dv
dr + \alpha^2 r^2 (dx^2+dy^2). \label{eq:me3}
\end{equation}
where $\beta^2 = ( 3 \alpha^2 - 8 \pi \rho_0) /3$ is a constant.
Because $\rho$ is the constant, the physical parameter can also be
trivially evaluated to give
\begin{equation}
\psi = \frac{1}{4\pi r^2} \dot{M(v)} \qquad \ p_r = p_{\bot} = -
\rho_0
\end{equation}
The results obtained suggest that constant string density star can
have negative pressure or tension, this is consistent with Katz and
Lynden Bell \cite{ck}.  This can happen particularly in the hybrid
star, which have ordinary positive pressure in the outermost layer
and tension in the interior.

 This is singular  star solution.  The reason
for calling singular can be seen by examining the various scalars
that can be built form Reimann tensor at $r=0$, e.g., the
Kretschmann for the metric (\ref{eq:de2}), has the form:
\begin{equation}\label{ks}
   \mbox {{\tt K }}= \frac{48 M^2(v)}{r^6} + \frac{512 \pi^2
   \rho^2}{3} + 24 \alpha^2 - 128 \alpha^2 \pi \rho
\end{equation}
So the Kretschmann scalar diverges along $r = 0$ for $M(v) \neq 0$,
establishing that metric (\ref{eq:me}) is scalar polynomial singular
\cite{he}. Proceeding as above the analysis of structure of this
singularity is initiated by study of the radial null geodesics that
satisfy
\begin{equation}
\frac{dv}{dr} = \frac{2 r}{\beta^2 r^3- M(v)} \label{eq:de2}
\end{equation}
The Eq. (\ref{eq:de2}) has a singular point at $r=0$ and $v=0$ and
we write
\begin{equation}
M(v) = \lambda v \; (\lambda > 0)  \label{eq:emv}
\end{equation}
Following the standard techniques one finds that the singularity is
center for all $\lambda$, no radial outgoing geodesic that will
terminate in the past at the singularity. Hence  the collapse will
always lead to a black hole in accordance with CCC.
\section{Concluding remarks}
In conclusion, we have given a class of non-spherical collapse
solutions (\ref{eq:me2}) and (\ref{eq:me3}), for two fluid, of the
Einstein field equations for the energy momentum tensor
(\ref{emsv}).  The solutions are asymptotically anti-de Sitter black
hole solutions ($1/2< k <1 $) with multiple horizon.  The general
metric (\ref{eq:me2}) depends on one parameter ($k$), two arbitrary
function of $v$ (subject to energy condition).  The class of the
solutions discussed above, in general, belongs to Type II fluid
defined in \cite{he}. When $m=m(r)$, we have $\psi$=0, and the
matter field degenerates to type I fluid \cite{ww}. We have used the
solutions to study string effects in the non-spherical null fluid
collapse.  It turns out that, non-spherical collapse of a two-fluid
in anti-de Sitter background, naked singularities do not form, in
accord with CCC.  On the other hand planar and cylindrical
black-holes form, giving explicit counter example of the hoop
conjecture. Note that this conjecture is not valid when $\alpha  =
0$.  This shows that the non-spherical collapse in anti-de Sitter
back ground may violate hoop conjecture but not CCC.

\end{document}